\documentstyle[12pt] {article}
\begin{document}
\input epsf
\parindent=15pt
\begin{center}
\vskip 1.5 truecm
{\bf CLASSICAL PION FIELDS IN THE PRESENCE OF SOURCE}\\
\vspace{1cm} M.G.Ryskin and A.G.Shuvaev\\ Theory Department, St.Petersburg
Nuclear Physics Institute\\ 188350, Gatchina, St.Petersburg, Russia.\\
e-mail: {\tt ryskin@thd.pnpi.spb.ru}\\
e-mail: {\tt shuvaev@thd.pnpi.spb.ru}
\end{center}
\begin{abstract}
Classical pion field similar to Disoriented Chiral
Condensate (DCC) is considered in the presence of the external source.
This field is similar to DCC in the sense that its isotopic orientation
is specified with a single vector at the whole space. We study the classical
field solutions in the nonlinear sigma-model both in the chiral limit with
massless pion and for the finite pion mass. In both cases the field resembles
the Coulomb field of charged particle however the nonlinear pion interactions
lead to the existence of several solutions. In the massless case and for the
very small size of the source there is the lot of classical solutions
with finite discrete energies. In the more realistic situation of large
nucleus (heavy ion) there are no stable solutions of the above type, but there
is the possibility for the formation of the quasistationary states.
They can live for a long time slowly decaying through the emission of very
soft pions. The structure and the energies of these solutions is investigated
numerically.
\end{abstract}
\vspace{1cm}

\section{Introduction}

The interaction of the soft pions at the comparatively low energies
is described by the effective Lagrangian
\begin{equation}
\label{Lpi}
L_\pi\,=\,\frac 12 \bigl[\,(\partial_\mu \sigma)^2\,+\,
(\partial_\mu \vec{\pi})^2\,\bigr],
\end{equation}
including three isovector pion fields $\pi_i$, $i=1,2,3$, and an auxiliary
scalar field $\sigma$ obeying the constraint
$$
\sigma^2\,+\,\vec{\pi}^2\,=\,f_\pi^2,
$$
where $f_\pi=92$~MeV is the pion decay constant.

The pion field represents the chiral phase of the quark condensate, that
is why the parameterization through the unitary matrix is quite natural
in this approach. Constructing the matrix
$$
U\,=\,\frac 1{f_\pi}\,\bigl(\,\sigma\,+\,i\vec{\pi}\vec{\tau}\,
\bigr), \qquad U^+U\,=\,1,
$$
the Lagrangian (\ref{Lpi}) takes the form
\begin{equation}
\label{Leff}
L_{eff}\,=\,\frac{f_\pi^2}4 {\rm Tr}\,\partial_\mu U\,\partial_\mu U^+.
\end{equation}

The Lagrangian (\ref{Leff}) involves non-linear terms responsible for the
many-pion interaction and allows for the classical solutions. The important
class of plane wave type solutions for the pion field was described by
A.A.Anselm \cite{1,2}, the Disoriented Chiral Condensate (DCC) being the
particular case of it with the wave vector $\vec k =0$ \cite{5, 6, 7, 8, 9}.

In the collision of high energy particles (or, better, heavy ions) the system
is "warmed up" to the temperatures at which the chiral symmetry is restored.
Then, in the course of cooling, the symmetry breaks again and the scalar,
$\langle \overline q q \rangle$, or pseudoscalar,
$\langle \overline q \gamma_5 \tau_a q \rangle$, condensate with the quantum
numbers of $\sigma$ or $\pi^a$ mesons settles out. All four orientation
(3 pions and $\sigma$-meson) are equivalent in the isotopic $O(4)$ space,
and if the pion condensate, $\langle \overline q \gamma_5 \tau_a q \rangle$,
is produced somewhere it means the formation of DCC domain.

Possible scenarios for the classical pion field (DCC) production at the high
energies and the experimental signatures of DCC have been widely discussed
in the literature \cite{5, 6, 7, 8, 9, 13, 18, 20, 21, 22, 23}.

One of the most significant evidence of DCC is the distribution over the
neutral and charged pions.

The multiplicity distribution obeys the Poisson law for independent
production, so the number of $\pi^0$ mesons is 1/3 of the total number of
pions in the large multiplicity event while the distribution of the ratio
$f=n^0/n_{tot}$ is close to $\delta(f-1/3)$ (here $n^0$ is the number of
$\pi^0$, $n_{tot}$ is the total number of pions).

In case of a classical pion field creation the orientation of the isotopic
vector $\vec A^a$, which determines the ratio of charged and neutral pions,
is chosen once for the whole classical field domain. It results into the
large probability,
\begin{equation}
\label{sqrtf}
\frac{dw}{df}\,=\,\frac 1{2 \sqrt f},
\end{equation}
of the events with anomalous small number of neutral pions \cite{an,1,5,7,8}.

There have not been still a real progress in experimental attempts
to observe such events (in $p \bar p$ collisions at FNAL collider
\cite{fnal} and in nucleus-nucleus reactions in CERN\cite{cern})
although the hope to create DCC in this conditions was not
originally much. The heavy ion collisions at the new colliders
(RHIC, LHC) seem to be more promising.

In this paper we deal with another possibility - with classical pion
fields, which can be produced and exist in the presence of an external
source i.e. nucleons or quarks of an incident nucleus. Contrary to
the Skyrmion type fields for which the isotopic direction of the vector
is strongly correlated with the coordinate space position (so called
"hedgehog ansatz") here the direction is described by the single vector
$\vec A^a$ at the whole space just as it is for Anselm solution or DCC.
Therefore the distribution (\ref{sqrtf}) for the ratio of charged and
neutral particles is still valid for the treated fields.

The structure and physical nature of these solutions is rather similar
to electromagnetic (Coulomb) field of a charged particle, however
the nonlinear pion interaction (\ref{Leff}) allows for the existence of
several solutions with different energies. The nucleon or nucleus in
the cloud of such fields can be regarded as an excited state or resonance.

Below we shall consider the spherically symmetric solutions at first in
the chiral limit, putting the pion mass $m_\pi=0$, and then we shall add
the mass term to the Lagrangian (\ref{Leff}).

The mass term looks like the source in the classical equation and gives
rise to the piece of the form $m_\pi^2 \sin \varphi$, where
$\varphi = |\pi|/f_\pi$.
The contribution of the external source is opposite in sign to
the mass term. For a weak fields $\varphi \ll 1$ the source acts as an
attractive potential and allows thereby for a stable
solution --  a stationary state. Even for a strong field $\varphi \sim  1$
the stationary state is possible in the case of large size source.
Once this classical solution is created in the heavy ion collision
(if created) it can live for the long time slowly decaying through
the emission of very soft pions (in the nucleus reference frame).

We describe the classical equation for the fields
surrounding the nucleon and possible excitations (resonances) of
this system in Section~2. The more interesting case of a large nucleus
is considered in Section~3.

\section{Classical solutions in chiral limit}

The pion interaction with fermions (quarks or nucleons) is given by the
microscopic Lagrangian
\begin{equation}
\label{Lsq}
L_f\,=\,\overline{q}\, i\gamma^\mu \partial_\mu q\,-\,
g\overline{q}\bigl(\sigma + i\vec{\pi}\vec{\tau}\gamma_5\bigr)q.
\end{equation}
Further we shall treat it as an interaction with external classical
source generally described by isoscalar
$\rho(x)=\langle \overline{q}(x) q(x)\rangle$ and isovector
$\rho_V(x)=\langle \overline{q}(x) \vec{\tau}\gamma_5 q(x)\rangle$
densities. For the simplicity we shall deal with the situation when
the source has zero isospin, so that $\rho_V(x)=0$, and the Lagrangian
takes the form
\begin{equation}
\label{Ls}
L\,=\,\frac{f_\pi^2}4 \,{\rm Tr}\,\partial_\mu U\,\partial_\mu U^+-
\frac 14 g f_\pi \rho \,{\rm Tr}\,\bigl(U\,+\,U^+\,-\,2\bigr),
\end{equation}
the last term being zero for $U=1$.

Upon varying the Lagrangian with respect to the pion matrix $U$,
the equation of motion yields
$$
\partial_\mu\,\bigl[U^+\partial_\mu U\bigr]\,=\,\frac g{2 f_\pi}\,
\rho(x)\,\bigl[U^+- U\bigr].
$$
In what follows the matrix $U(x)$ is sought among the kind of exact
solutions found in refs.\cite{1,2}:
\begin{equation}
\label{ans}
U(x)\,=\,V^{-1}e^{i\tau_3 f(x)}\,V,
\end{equation}
where $V$ is an arbitrary but constant unitary matrix. The function
$f(x)$ obeys the equation
$$
\partial^2 f(t,x)\,=\,\frac g{f_\pi}\,\rho(x)\,\sin\, f(t,x).
$$

We are interested in the stationary solutions for which
$$
\Delta f(x)\,=\,\frac g{f_\pi}\,\rho(x)\,\sin\, f(x),
$$
where $\Delta = -\partial_i^2$ is the Laplace operator.
Supposing the function $f(x)$ to be decreasing at the space
infinity this equation can be transformed to the nonlinear
integral form
\begin{equation}
\label{int}
f(x)\,=\,\frac 1{4\pi}\frac g{f_\pi}\int\frac{d^3y\,\rho(y)}{|x-y|}\,
\sin \,f(y).
\end{equation}

The total energy accumulated in the pion field is generally finite:
\begin{equation}
\label{E}
E\,=\,gf_\pi\,\int d^3x\,\rho(x)\bigl[\frac 12 f\,\sin \,f\,+
\,\cos \,f\,-\,1\bigr].
\end{equation}

There are two different regimes for the equation (\ref{E}) - the regime
of small and large size of the source.

In the first case the density is concentrated in the small space, and
since we are looking for a smooth function we can put $\sin f(y)
\approx \sin f(0)$ in the integrand (\ref{int}). Then we obtain the
Coulomb-like potential at the finite distances from the source
$$
f(x)\,=\,\frac 1{4\pi}\frac g{f_\pi}\frac 1{|x|}\sin\,f,
$$
with the charge-like constant $f=f(0)$ obeying the equation
\begin{equation}
\label{fsinf}
f\,=\,\frac 1a \sin\,f,
\end{equation}
in which the dimensionless parameter $a$,
$$
\frac 1a\,=\,\frac g{4\pi f_\pi}\,\int\,d^3y\,\frac{\rho(y)}{|y|},
$$
is proportional to the radius of the source.

The equation (\ref{fsinf}) has a set of solution for $a \ll 1$
$$
\sin\,f_n\,\simeq \pi a n,\quad n\,=\,0,\pm 1,\pm 2,\ldots,\quad
|n|\le \frac 1{\pi a}.
$$
As a result the pion energy (\ref{E}) turns out to be quantized
despite the pure classical treatment:
\begin{equation}
\label{En}
E_n\,=\,gf_\pi\bigl[\frac 12 f_n\,\sin \,f_n\,+
\,\cos f_n\,-\,1\bigr].
\end{equation}

The lesser the effective dimensionless radius of the source, $a$,
the more energy levels appear. For $a\to 0$ they form two
quasi-continuous bands for even and odd states approximated
for not very large $n$ as
\begin{equation}
\label{band}
E_n\,\simeq\,gf_\pi\bigl[(-1)^n-\,1\,+\,\frac 12 \pi^2 n^2 a \bigr],
\end{equation}
with energy spacing $\sim g \pi^2 f_\pi a$ between the levels in the
each band.

The lower levels in the odd band in (\ref{band}) have the negative
energy. One has to emphasize here that $E_n$ are only the pion field
energies whereas the total energy comprises, except this, the energy
of the source itself, that is the energy of the quarks or nucleons
inside. The smaller the volume they occupy the large energy they have.
Therefore there is a competition between these two effects which,
could lead under certain conditions to a bound state formation.

In the second regime there are no finite energy solutions except
for the $f=0$. Indeed, after rescaling the variables $\rho(x)=
1/r_0^3 \overline \rho(x/r_0)$, $f(x)=\overline f(x/r_0)$, where
$r_0$ is a characteristic size, where the density is non zero,
the equation (\ref{int}) reads
$$
\overline f(z)\,=\,\frac g{4\pi f_\pi
r_0}\int\frac{d^3z^\prime\,\overline\rho(z^\prime)}{|z-z^\prime|}\,
\sin\,\overline f(z^\prime),
$$
so only the zero solution is allowed for large enough $r_0$.

The simple estimation shows that the critical radius $r_0$ needed
for the nontrivial solution should be smaller than 1~fm in the order of
magnitude. The energy $E_1$ for $r_0\sim 1$~fm is of the order of the mass
difference between the baryon resonances and it is not excluded that some/part
of this states are due to an excitation of the "Coulomb-like" pion field
considered above.

\section{Large size source}

The last statement on the absence of the solution for the large source
is valid only for the fixed constant $g$ while the
value of $g=g_{\pi NN}\cdot A$ increases for a heavy ion with the atomic
number $A$ faster than the characteristic radius $r_0\propto A^{1/3}$.
This is an interesting case which should be discussed in more detail.
Let us put the nucleon density $\rho =const$ inside the nuclear and
include the pion mass. It results into the equation
\begin{equation}
\label{n1}
\Delta f(x)\,=\,\left(\frac g{f_\pi}\,\rho(x)\,-m^2_\pi\right)\sin\, f(x),
\end{equation}
where the density $\rho(x)=\rho_0\theta(R-|x|)$ and $R$ is the nuclear radius.
For the normal nuclear density the effective strength of the source
$G=\rho_0\frac g{f_\pi}\simeq 4.7$ fm$^{-2}$.

For the weak fields the equation can be linearized and takes the simple form
\begin{equation}
\label{n2}
\Delta f(x)\,=\, (G-m^2_\pi) f(x).
\end{equation}
The well known spherically symmetric solution reads
$$
f(r)\,=\, B_1\frac{\sin (rb)}{r}\;\;\;\;\; \mbox{for $r=|x|<R$}\; ,
$$
with $b=\sqrt{G-m^2_\pi}$ and
\begin{equation}
\label{n3}
f(r)\, = \, B_2\frac{\exp (-rm_\pi)}{r}\;\;\;\;\; \mbox{for $r>R$}\; .
\end{equation}
In order to match the solutions at $r=R$ the logarithmic derivative of the
first expression should obey the condition
\begin{equation}
\label{n4}
\frac{d\ln(rf(r))}{dr}_{(\mbox{at $r=|x|=R$})}\,
=\, b\cdot ctg (Rb)\, =\, -m_\pi \; .
\end{equation}
The fine tuning of nuclear {\it radius} $R$, which is needed to satisfy
(\ref{n4}), looks, of course, rather unlikely. However it is possible
to provide the matching in non-linear case by choosing
the value of $f(0)\sim O(1)$ (for large enough $b>\pi/R$). We find such
solutions numerically for a reasonable ion radius $R=5.6$ - $5.9$~fm with the
field amplitudes $f(0)\simeq 1$ - $2$, respectively. The results are shown in
Fig.1 by the solid curves.

This solution in some sense has a similar nature with the
classical Coulomb field, the amplitude of which is fixed by the
charge (and the size) of the source as well.

To calculate the energy of the pion field we include the pion mass term
$m^2_\pi$ in (\ref{E}) and obtain a very small value $E=-8$~MeV
for $f(0)=0.96$ (this is still a weak field) but $E=-280$~MeV for the
case of $f(0)=2$.

The fields shown by the solid curves in Fig.1 look as the result of
the pion condensation in a heavy nuclear which was studied many years
ago \cite{Mig} mainly in terms of the Fermy Liquid Theory.
A possibility to observe the pion condensation in heavy ion collisions
is discussed in \cite{Pir}.

Unfortunately, such a solution can not be realized.
There is no pion condensation
at the normal nuclear density $\rho_0$. What is happened?

The Lagrangian (\ref{Lsq}) is correct being written in terms of
the quarks and pions. We have applied it to the nucleon instead
of the quarks.
Taking the value $g=g_{\pi NN}$ we assume the whole nucleon mass
to be generated by the interaction of the nucleon with the classical
$\sigma$ field $<\sigma >=f_\pi$.
In the framework of the effective Lagrangian (\ref{Lpi}),(\ref{Leff})
it leads to a very strong interaction of the soft pions with the
nucleons. On the other hand the $\pi N$ scattering amplitude is
rather small corresponding to a so-called $\sigma$-term, which is
about 20-30 times less than the proton mass.

That is why we take a smaller value of the
coupling $g$ (say, $G=0.18$~fm$^{-2}$) in the better agreement
with the nucleon $\sigma$-term. Then a heavy ion acts as a rather
weak source.
For $G<0.067$~fm$^{-2}$ and a reasonable ion size ($R\sim 6$~fm)
the resulting attractive potential is even not sufficiently strong
to form a bound state.

However it is possible to form for $G=0.18$~fm$^{-2}$ the quasistationary
long-living state, which decays only due to the non-linear effects.
Indeed, in a weak field approximation one can find the time dependent
solution in the form $f(t,x)=exp(iEt)f(x)$, where the function $f(x)$
satisfies eq.(\ref{n2}) with $G-m^2_\pi$ replaced by $G+E^2-m^2_\pi$.
Choosing the eigenenergy $E\simeq 130$~MeV to provide
the matching at $r=R$ we obtain the solution shown in Fig.1 by the dotted
curve (for $R=5.9$~fm).

Being produced in a heavy ion collision (if it is)
it can live for the long time slowly decaying through the emission of very
soft pions (in the nucleus reference frame) due to non-linear terms in the
Lagrangian (\ref{Leff}). The life-time of the solution decreases with the
field amplitude $f$, which determines the multiplicity $n_\pi$ of the pions
in the classical field. The dotted curve presented in Fig.1 corresponds to
$n_\pi =2E\, f^2_\pi \cdot\int f^2(x)d^3x\simeq 12.3$

The existence of such a solution may help in production of
a "breather" like classical pion fields \cite{GBK}. Interaction with the
source (nuclear matter) makes this solution more stable and long-lived.
The presence of the source could also stabilize under certain conditions
the solutions of "pion string" kind \cite{Zhang}, for which the direction 
of the isotopic vector $\vec A^a$ is correlated with the azimuthal angle 
in the coordinate $x,y$ plane.

Nevertheless the production of the classical pion field we discussed
first (solid curves in Fig.1) can not be completely excluded
in the high energy heavy ion collisions. If a kind of "quark bag"
(large drop of quark plasma) would be formed after collision instead of
the outgoing "nuclear", then we, probably, come back to the {\it quark}-pion
Lagrangian (\ref{Lsq}) with a large coupling $g$ and the corresponding
value $G\sim 3 - 5$~fm$^{-2}$. The interaction now would be enough to
form a stationary classical pion field (pion condensate) in the drop of
quark matter.
\smallskip

\noindent
{\bf Acknowledgments}\\
One of us (A.S.) is grateful to Prof. K.~Goeke for hospitality
and to Prof. L.~McLerran for discussions.\\
This work is supported by grants DFG-RFFI-436-M3/540 and RFBR 98-02-17636.
\begin{figure}[t]
\epsfxsize=15cm
\epsfysize=15cm
\vspace*{1cm}
\centering
\epsffile{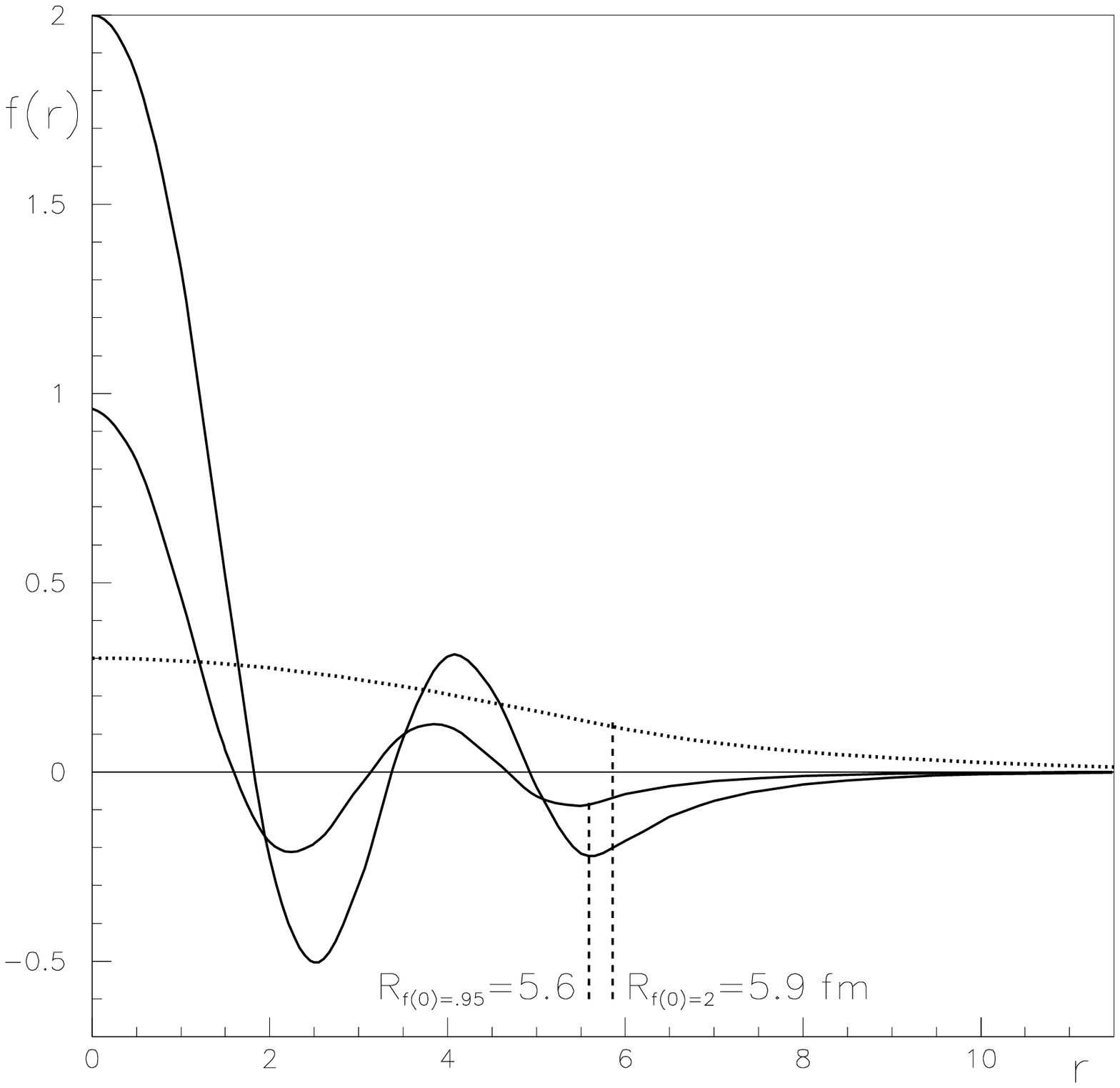}
\caption{Amplitude of the classical pion fied in the presence of
a large size nucleon or quark source (heavy ion) versus the radius.
Radius of the ion $R$ is marked by the dashed lines. Solid lines - strong
coupling $G=4.7$ fm$^{-2}$, dotted line - weak coupling $G=0.18$ fm$^{-2}$.}
\end{figure}


\begin{thebibliography}{99}
\bibitem{1} A.A.Anselm, Phys.Lett. {\bf B217} (1989) 169.
\bibitem{2} A.A.Anselm, M.Bander, JETP.Lett. {\bf 59} (1994) 503.
\bibitem{5} J.D.Bjorken, Int.J.Mod.Phys. {\bf A7} (1992)
4189; Acta Phys.Pol. {\bf B23} (1992) 561.
\bibitem{6} J.-P.Blaizot, A.Krzywicki, Phys.Rev. {\bf D46} (1992) 246.
\bibitem{7} K.L.Kowalski, C.C.Taylor, CWRUTH-92-6, hep-ph/9211289.
\bibitem{8} J.D.Bjorken, K.L.Kowalski, C.C.Taylor, SLAC-PUB-6109
(1993).
\bibitem{9} K.Rajagopal, F.Wilczek, Nucl.Phys. {\bf B399} (1992) 395.
\bibitem{13} J.D.Bjorken et al., MiniMax: A Revised Proposal for
T-864, April 1993.
\bibitem{18} P.V.Ruuskanen, Z.Phys. {\bf C38} (1988) 219; Acta
Phys.Pol. {\bf B18} (1987) 551.
\bibitem{20} S.Gavin, A.Gocksch, R.D.Pisarski, Phys.Rev.Lett. {\bf
72} (1994) 2143.
\bibitem{21} S.Gavin, B.M\"uller, Phys.Lett. {\bf B329} (1994) 486.
\bibitem{22} Ian.I.Kogan, JETP Lett. {\bf 59} (1994) 307.
\bibitem{23} A.A.Anselm, M.G.Ryskin, A.G.Shuvaev. Problems of high energy
physics, International winter school on theoretical high energy
physics, PNPI, St.Petersburg, 1995; Z.Phys {\bf A354} (1996) 333.
\bibitem{an} I.V.Andreev, JETP Lett. {\bf 33} (1981) 367.
\bibitem{fnal} MiniMax Collab.: T.C.Brooks et al., hep-ex/9906026.
\bibitem{cern} WA98 Collab.: M.M.Aggarwal et al., Phys. Lett. {\bf B420}
(1998) 169.
\bibitem{Mig} A.B.Migdal, Rev. Mod. Phys. {\bf 50} (1978) 107,\\
A.B.Migdal, E.E.Saperstein, M.A.Troitskii, D.N.Voskresensky,
Phys. Rep. {\bf 192} (1990) 179.
\bibitem{Pir} H.J.Pirner, D.N.Voskresensky, Phys. Lett. {\bf B343} (1995) 25.
\bibitem{GBK} V.A.Gani, A.E.Kudryavtsev, T.I. Belova, B.L.Druzhinin,
Phys. At. Nucl. {\bf 62} (1999) 895.
\bibitem{Zhang} X.Zhang, T.Huang and R.H.Brandenberger, Phys. Rev. {\bf D58}
(1998) 027702; hep-ph/9711452.

\end{thebibliography}
\end{document}